\documentclass{article}

\PassOptionsToPackage{numbers,square,sort&compress}{natbib}
\usepackage[preprint]{neurips_2024}
\usepackage[T1]{fontenc}
\usepackage[utf8]{inputenc}
\usepackage{microtype}
\usepackage{amsmath, amssymb}
\usepackage{booktabs}
\usepackage{float}
\usepackage{graphicx}
\usepackage{xcolor}
\usepackage{enumitem}
\usepackage{titlesec}
\usepackage{hyperref}
\usepackage{url}

\hypersetup{
  colorlinks=true,
  linkcolor=blue,
  citecolor=blue,
  urlcolor=blue,
  pdftitle={Inherited Circuits, Learned Semantics: How Fine-Tuning Creates Evasion Vulnerabilities Invisible to Standard Evaluation}
}

\titlespacing*{\section}{0pt}{1.0ex plus 0.2ex minus 0.1ex}{0.45ex}
\titlespacing*{\subsection}{0pt}{0.8ex plus 0.2ex minus 0.1ex}{0.35ex}
\title{Inherited Circuits, Learned Semantics: How Fine-Tuning Creates Evasion Vulnerabilities Invisible to Standard Evaluation}
\author{%
  Ryan Fetterman \\
  Cisco Talos\\
   \texttt{rfetterm@cisco.com}
}
\date{}

\begin{document}
\raggedbottom

\maketitle
\begin{abstract}
LLMs are increasingly fine-tuned for specific security classification tasks, and the resulting models are typically evaluated on held-out examples from the same distribution as their training data. We show that this evaluation practice misses a class of vulnerability introduced by fine-tuning itself: a model trained on a corpus where specific indicator tokens are strongly correlated with maliciousness can learn token-level semantics that preserve canonical accuracy while failing under behavior-preserving transformations. In our PowerShell setting, these transformations include alias substitution, command reconstruction, string construction, execution indirection, and case mutation.

To understand the mechanism, we study a natural base/fine-tuned model pair, Llama-3.1-8B-Instruct and Foundation-Sec-8B-Instruct, on matched PowerShell classification cohorts. Using causal interventions, we localize the classification circuit to a small late-attention bundle and show that this route is inherited from Llama rather than created by fine-tuning. Fine-tuning concentrates and semantically specializes this inherited structure, adding associations between canonical command indicators and malicious classification. Those associations improve baseline classification behavior but create transformation-sensitive attack surfaces: a three-tier evasion benchmark finds Foundation-Sec misses across multiple transformation categories that Llama does not share.

We derive a practical pre-deployment monitoring method that combines a linear probe from base-model activations at the classification boundary with an indicator-token sign test to cheaply track post-fine-tuning drift, identify command families where canonical indicators change role, and prioritize targets for red-team variant generation. More broadly, these results caution against treating small task-specific fine-tunes as straightforwardly safer security classifiers: specialization can convert inherited model structure into brittle indicator rules that preserve held-out accuracy while expanding the evasion surface. Robust AI-enabled security will require specifying the full transformation space of the task and monitoring semantic drift through fine-tuning.
\end{abstract}


\section{Introduction}
Teams fine-tuning LLMs for security classification typically evaluate the resulting model on held-out examples drawn from the same distribution as the training data. This measures accuracy on canonical inputs but does not reveal whether the model has become reliant on specific indicator tokens in ways that make it vulnerable to behavior-preserving transformations. From an attacker's perspective, these transformations are cheap and common in PowerShell: a script can replace \texttt{Invoke-WebRequest} with \texttt{iwr}, reconstruct \texttt{Invoke-Expression} through a format string, or move execution through a call operator while preserving malicious behavior. From a defender's perspective, held-out canonical accuracy does not test this attack surface. A model that learned canonical indicator semantics can classify the canonical form correctly, and will appear fine on held-out test data, while silently failing on transformed scripts.

This paper addresses that gap with a pre-deployment detection method. The core insight is that transformation-sensitive evasion has a detectable signature in the model's internal representations, observable on canonical clean inputs before any evasion variants are generated. When fine-tuning causes an indicator token to shift from driving malicious classification to suppressing it because the model has learned to classify that family through surrounding payload context rather than the indicator itself, that family has become a transformation attack surface. This shift is directly measurable: ablate the indicator tokens from canonical malicious scripts, compare the change in confidence between the base model and the fine-tuned model, and flag families where the sign diverges. A family where the fine-tuned model's confidence increases when indicator tokens are removed (suppressor role) while the base model's confidence decreases (driver role) is the target for transformation-focused red-teaming.

We validate this method on Foundation-Sec-8B-Instruct and its base model, Llama-3.1-8B-Instruct, using a 293-pair matched cohort of PowerShell scripts and a three-tier evasion benchmark. The indicator-token sign test identifies \texttt{Invoke-WebRequest} as the family with the clearest model divergence, and the benchmark confirms the targeted risk with consistent misses on \texttt{iwr} alias substitution. The benchmark also finds transformation-sensitive failures on \texttt{Invoke-Expression} format-string reconstruction, alternating-case \texttt{Invoke-Expression}, and alternating-case \texttt{IEX} alias variants. Llama has zero misses on the same evaluated variant sets. This is the central tradeoff: Foundation-Sec improves baseline classification behavior but loses robustness that Llama retains. The sign test result, computed entirely on canonical inputs, correctly anticipates the family-level \texttt{Invoke-WebRequest} outcome before any evasion variant is attempted.

The contributions of this paper are:
\begin{itemize}[leftmargin=1.5em,itemsep=0.15em,topsep=0.15em,parsep=0pt,partopsep=0pt]
  \item \textbf{A mechanistic account of security fine-tuning:} causal interventions show that Foundation-Sec inherits a Layer-12 classification route from Llama rather than creating a new PowerShell detector. Fine-tuning concentrates and semantically specializes this route, producing signal reversal in successful evasion cases rather than simple route deletion.
  \item \textbf{A security failure mode for fine-tuned classifiers:} Foundation-Sec improves baseline classification behavior but develops transformation-sensitive misses that Llama does not share. A three-tier evasion benchmark confirms failures across direct rewrites, command reconstruction, and case mutation, and prompt-level remediation is uneven: fixing one family can open or amplify misses in another.
  \item \textbf{A reusable evasion benchmark artifact:} the benchmark is built from matched seed/variant pairs, syntax-preserving rewrites, parse or manual-validation checks, and behavioral invariants. This lets other teams attribute failures to specific transformation classes rather than treating obfuscation as a generic leaderboard.
  \item \textbf{A pre-deployment monitoring method:} a linear probe at the classification boundary flags family-level representation drift, and an indicator-token sign test identifies families where canonical indicators shifted from driver to suppressor role. The output is a ranked list of command families to red-team, not a per-script evasion prediction.
\end{itemize}

Code, benchmark artifacts, and public figures are available at \href{https://github.com/fetterm4n/inherited-circuits}{\texttt{fetterm4n/inherited-circuits}}.

\section{Prior Work and Background}
A body of research has established that transformer-based language models can often be understood in terms of small, identifiable internal components, such as attention heads and feed-forward layers, that causally drive specific behaviors \citep{cammarata2020thread,elhage2021mathematical}. Studies have shown these components can be identified through causal interventions: selectively disabling or substituting internal activations and measuring how the model's output changes \citep{wang2023interpretability,conmy2023automated}. More recent work has extended these tools toward automated discovery and toward understanding how fine-tuning reshapes these internal structures \citep{ameisen2025attributiongraphs,wang2025finetuningcircuits}. A consistent finding is that fine-tuning tends to modify how existing components are weighted and used rather than creating entirely new structures from scratch, which is directly relevant to interpreting our Foundation-Sec results.

The model lineage matters for interpretation. Foundation-Sec-8B was built from the Llama 3.1 8B base architecture through continued pretraining on a curated cybersecurity corpus \citep{kassianik2025foundationsecbase}. Foundation-Sec-8B-Instruct extends this with instruction tuning for chat-style security use \citep{weerawardhena2025foundationsecinstruct}. The model was previously evaluated favorably in PowerShell classification, with a reported \(+16\%\) precision gain over the baseline in a threat-hunting workflow \citep{fetterman2025foundationsecpowershell}. Because Foundation-Sec and Llama share the same underlying architecture, any differences in their internal classification behavior can be attributed to fine-tuning rather than to architectural changes, making this a clean comparison for isolating what security fine-tuning adds or changes.

The evasion side of this paper connects to a broader pattern in security research: detectors that perform well on canonical inputs can fail under functionality-preserving rewrites that fall outside the training distribution \citep{lucas2023adversarial,mukherjee2023evading}. The specific concern here is that LLM-based classifiers may rely on brittle surface indicators. This is the same concern, applied to a model whose internal mechanism can now be examined directly.

A transformer processes text by passing it through a sequence of \emph{layers}. Each layer contains two types of components: \emph{attention heads}, which look back at earlier parts of the input and pull in relevant information, and \emph{MLP blocks} (feed-forward sublayers), which apply learned transformations to the current state. All components write their outputs into a shared running representation called the \emph{residual stream}, which accumulates information across layers and ultimately determines the model's output. The notation \texttt{L12H15} means Layer 12, Attention Head 15. \texttt{resid\_pre13} refers to the state of the residual stream immediately before Layer 13, the point we use as the classification boundary in this study.

We use the term \emph{circuit} to mean a small subset of model components, specifically attention heads and MLP blocks, whose outputs causally determine the model's classification decision for a given input type. Identifying a circuit means showing that these components are both necessary (removing them weakens the decision) and sufficient (copying their activations from a malicious example into a benign context recreates the malicious classification). A \emph{route} is one specific pathway through that circuit, and a \emph{late-stage bundle} is the group of late-layer heads that carry the strongest classification signal near the output.

Two intervention techniques drive the analysis. \emph{Path patching} tests sufficiency: copy the internal activations of a candidate route from a malicious script into the same positions in a benign script, and measure how much the classification score shifts toward malicious. \emph{Head ablation} tests necessity: disable a candidate group and measure how much the malicious classification weakens. When we say a route is \emph{validated}, we mean it passes both tests on the 293-pair matched cohort. When we say \emph{signal inversion}, we mean the model still internally generates evidence of maliciousness, but a later component reverses how that evidence affects the final output.

\section{Problem Setting}
The task is binary malicious-code classification for PowerShell scripts. The key mechanistic difficulty is that many suspicious strings are not uniquely malicious: benign administrative scripts can contain tokens such as \texttt{IEX}, \texttt{FromBase64String}, \texttt{DownloadString}, \texttt{Invoke-WebRequest}, or \texttt{-EncodedCommand}. We therefore use matched benign and malicious examples sharing suspicious indicators, so that the interpretability problem is not trivially solved by superficial token presence alone. Appendix~\ref{app:dataset-cohorts} describes the source corpus, length controls, and matched evaluation cohorts used to construct the 293-pair intervention set.

\section{Methods}
Appendix~\ref{app:claim-ladder} summarizes how discovery, validation, comparison, and evasion claims are separated in the current study.

\subsection{Discovery}
The goal of the discovery phase is to identify a small number of candidate components that repeatedly carry malicious-versus-benign evidence before we run the stronger causal tests below. We first examine which attention heads consistently focus on suspicious command spans, URLs, or execution sites across matched script pairs. We then run coarse ablations that disable individual heads or entire layer bands to ask whether removing a candidate weakens the malicious classification enough to justify studying it further \citep{cammarata2020thread,elhage2021mathematical,conmy2023automated,ameisen2025attributiongraphs}.

Two discovery techniques are especially important. First, we define a \emph{malicious direction} in the model's internal representation, a vector pointing from how the model represents benign scripts toward how it represents malicious ones, and test whether specific internal sites are pushing the model along that direction. This distinguishes components that are genuinely carrying classification-relevant evidence from those that are merely active. Second, we decompose the contribution at each site into individual head outputs, which lets us identify a small set of dominant heads rather than treating an entire layer as a black box. These analyses narrow the search space so that the final causal claim can be stated precisely rather than vaguely attributed to ``early'' or ``late'' layers.

\subsection{Validation}
The final claim is based on grouped interventions from the 293-pair cohort:
\begin{itemize}[leftmargin=1.5em,itemsep=0.2em,topsep=0pt,parsep=0pt,partopsep=0pt]
  \item \textbf{Grouped path patching} is used primarily as a sufficiency test.
  \item \textbf{Grouped head ablation} is used primarily as a necessity test.
\end{itemize}

In simple terms, a \emph{sufficiency} test asks whether a candidate route is enough to recreate the relevant behavior when its activations are patched in from a malicious example. A \emph{necessity} test asks whether the behavior weakens when that route is removed or disabled. These are related but not identical questions: a route can be sufficient without being strictly necessary if the model has backup pathways, and it can be necessary in a narrow setting without being the cleanest transportable route for patching.

Throughout the paper, \emph{mean intervention effect} refers to the average shift produced by an intervention on the model's malicious-versus-benign decision score, measured as the post-intervention score minus the original score. A more negative mean effect therefore indicates that the intervention weakens the malicious judgment; for patching and ablation experiments that remove or corrupt the route, larger negative values indicate stronger causal dependence on the removed route. \emph{Flip rate} reports how often that intervention is strong enough to change the model's final predicted label.

\subsection{Reusable Evasion Benchmark Construction}
We build the evasion benchmark as a reusable seed/variant attribution harness rather than a broad obfuscation leaderboard. Candidate variants are generated only from malicious seed scripts that the model initially classifies correctly, so the benchmark measures failures relative to seeds that already pass the baseline classifier. Each accepted row is a matched pair: the canonical seed and a behavior-preserving rewrite that changes a specific syntactic indicator while preserving the malicious operation. Throughout this section, \texttt{evasion} means behavioral evasion of classification, that is, a malicious variant that the model now predicts benign. The central mechanistic question is whether successful evasion removes the validated malicious-evidence route or instead preserves that route while later computation changes how that evidence is used. In the best-localized current case, this change appears at the Layer 13 boundary: feed-forward computation before that boundary introduces a competing signal that reverses the effect of the surviving late-stage bundle.

The implemented benchmark currently instantiates three reusable tiers across four transformation categories: \texttt{keyword\_hiding}, \texttt{string\_construction}, \texttt{execution\_indirection}, and \texttt{case\_mutation}. The first tier  (\texttt{direct\_v1}), relies on alias substitution, call-operator indirection, token splitting, and literal or method-name reconstruction. Tier 2, (\texttt{reconstructive\_v2}) stresses the classifier with character-level and command-reconstruction patterns that are often seen in obfuscated PowerShell, including method-name reconstruction, format-string reconstruction, Base64 or ASCII recovery, subexpression strings, zero-width normalization, and alternate quoting or backticks. The third tier (\texttt{case\_mutation\_v3}) applies lowercase and alternating-case variants while preserving the command's syntactic identity, position, and arguments. These techniques are chosen because they preserve the operational behavior while perturbing the lexical indicators a security classifier might over-weight.

\begin{table}[!htbp]
\centering
\caption{Reusable evasion benchmark structure. The \texttt{direct\_v1} tier applies direct syntax-preserving rewrites to visible command, token, and method forms, \texttt{reconstructive\_v2} uses composed expressions that reconstruct suspicious commands or strings at runtime, and \texttt{case\_mutation\_v3} changes token casing while preserving syntax and arguments. Variants are accepted only after parse or manual-validation checks and invariant checks for valid PowerShell syntax and preserved behavior.}
\label{tab:evasion-methods}
\begin{tabular}{p{0.03\linewidth}p{0.18\linewidth}p{0.44\linewidth}p{0.3\linewidth}}
\toprule
Tier & Family & Representative rewrite & Current role \\
\midrule
\texttt{v1}& \texttt{keyword hiding}& \texttt{Invoke-WebRequest ...} $\rightarrow$ \texttt{iwr ...} & Direct rewrite of visible command tokens.\\
\texttt{v1}& \texttt{execution indirection}& \texttt{iex \$x} $\rightarrow$ \texttt{\& ([scriptblock]::Create(\$x))} & Preserve execution while replacing direct invocation with call-operator indirection. \\
\texttt{v1}& \texttt{string construction}& \texttt{DownloadString(...)} $\rightarrow$ \texttt{PSObject.Methods ['Download'+'String'].Invoke(...)}& Direct method-name reconstruction without changing payload behavior.\\
\texttt{v2}& \texttt{keyword hiding}& \texttt{Invoke-Expression} $\rightarrow$ \texttt{\&(('{0}{1}' -f 'Invoke-','Expression'))} & Runtime command reconstruction through format strings or related expressions.\\
\texttt{v2}& \texttt{string construction}& \texttt{Invoke-WebRequest} $\rightarrow$ Base64/ASCII, subexpression, or zero-width-normalized reconstruction & Runtime recovery of suspicious strings or command forms.\\
\texttt{v3}& \texttt{case mutation}& \texttt{Invoke-Expression} $\rightarrow$ \texttt{InVoKe-ExPrEsSiOn} & Preserve command identity and arguments while changing the token embedding surface.\\
\bottomrule
\end{tabular}
\end{table}

The implemented benchmark pipeline is more specific than a generic ``obfuscation sweep.'' It first builds a malicious seed manifest restricted to baseline-correct malicious scripts. It then applies only a fixed set of allowed rewrites from the requested preset (\texttt{direct\_v1} or \texttt{reconstructive\_v2}), and each rewrite is used only on scripts where it makes sense. For example, a rewrite that hides \texttt{Invoke-WebRequest} is only applied to seeds that actually contain that command, and a rewrite intended for cross-runtime PowerShell is only used when the target script fits that runtime setting. Each generated variant is then reviewed with static parsing, using the PowerShell parser when the required runtime is available and a \texttt{tree-sitter} fallback otherwise, together with explicit invariant checks to ensure underlying behavior is not changed. The invariant checks require preservation of URL sets, executable or path-like literals, and \texttt{-EncodedCommand} equivalence where relevant, and they also enforce technique-specific argument-preservation checks such as preserved request arguments or preserved process-launch arguments. The malicious seed / variant pairing is what makes the mechanistic experiments interpretable: we measure the same circuit on the same script content under two conditions (literal token present vs. absent) and attribute any difference in the residual stream directly to the token transformation. Another team can adapt the artifact to a new command family by defining seed filters, allowed behavior-preserving rewrites, parse or runtime requirements, and technique-specific invariants for the behavior that must remain fixed.

\section{Evaluation \& Results}
This section reports the empirical evidence for the paper's three linked claims: the classifier route is inherited, fine-tuning changes how that route is used, and those changes create measurable evasion risk. We begin with causal circuit results, then compare Foundation-Sec against its Llama base model to separate inherited structure from fine-tuning-specific semantics. We then evaluate the operational consequence with a three-tier evasion benchmark covering direct rewrites, command reconstruction, and case mutation. Finally, we use the same activation site to derive a family-level monitoring signal for pre-deployment red-teaming.

\subsection{Circuit Results}
\subsubsection{Validated Late Route}
On the 293-pair cohort, causal patching supports a minimal direct branch:
\[
\texttt{L0H11} \rightarrow \texttt{L12H15/L12H5/L12H4}
\]
The mean patching effect is approximately \(-3.614\), which means that the intervention reduces the model's average malicious-versus-benign decision margin from about \(4.71\) to about \(1.10\). In other words, it removes roughly \(77\%\) of the average malicious decision margin and flips the final prediction on \(87/293\) pairs (\(0.297\)).

The stronger late-stage bundle is:
\[
\texttt{L12H15/L12H5/L12H4/L12H28}
\]
Its mean patching effect is approximately \(-3.824\), which reduces the same average malicious-versus-benign decision margin from about \(4.71\) to about \(0.89\). In other words, it removes roughly \(81\%\) of the average malicious decision margin and also flips the final prediction on \(87/293\) pairs (\(0.297\)).

The role of \texttt{L12H2} is more nuanced. Removing \texttt{H2} improves the main path-patching result, while the full top-five bundle remains slightly stronger under grouped ablation. Our current interpretation is therefore that \texttt{L12H2} is an auxiliary family-sensitive helper, not part of the most stable sufficiency core.

\subsubsection{Interpretation of Early and Late Stages}
The results support an early Layer 0 detector family, especially \texttt{L0H11} and \texttt{L0H9}. However, the final validated claim in this draft does not rest on a broad early-head story. Instead, it uses \texttt{L0H11} as the clearest entry point into the validated late route. The stronger late-stage evidence localizes to attention around Layers 12--13 and to feed-forward computation before the Layer 13 boundary that becomes especially important in the successful-evasion setting.

\subsubsection{Family-Level Variation}
The 293-pair cohort also reveals a useful family-level pattern. The best patch route is not identical across all indicator families. The minimal branch is strongest for \texttt{DownloadFile}, the fuller top-five late bundle is strongest for \texttt{DownloadString} and \texttt{IEX}, and the \texttt{H2}-free carrier (late Layer-12 head bundle with L12H2 removed) is strongest for \texttt{-EncodedCommand}, \texttt{Invoke-Expression}, and \texttt{Invoke-WebRequest}. This pattern supports the current interpretation that \texttt{H2} is not part of the most stable sufficiency core. One caveat: the \texttt{-EncodedCommand} family (7 pairs) shows zero flip rate under all patching variants, indicating that the circuit claim is weakest for this family and that it warrants a cautionary note; the other six families show consistent negative mean deltas across all route variants.

At the same time, grouped ablation suggests that \texttt{H2} still contributes a necessity signal across most families: the top-5 bundle (mean \(\Delta = -2.174\)) is stronger than the H2-free carrier (mean \(\Delta = -1.893\)), confirming \texttt{H2} as an auxiliary helper that matters for necessity even while it weakens patching sufficiency. The family-level picture therefore reinforces the write-up distinction between a stable late-stage bundle and a narrower auxiliary helper.

\begin{table}[!htbp]
\centering
\caption{Family-by-family patching results across the four main late-route variants on the 293-pair cohort. Each route cell reports \texttt{mean $\Delta$ / flip rate}, where a more negative \texttt{mean $\Delta$} indicates a stronger weakening of the malicious judgment under the intervention and therefore stronger dependence on the intervened route. Route abbreviations are: Minimal = \texttt{L0H11 $\rightarrow$ L12H15/L12H5/L12H4}, Top-5 = \texttt{L12H15/L12H5/L12H4/L12H2/L12H28}, H2-free = \texttt{L12H15/L12H5/L12H4/L12H28}, and H28-free = \texttt{L12H15/L12H5/L12H4/L12H2}.}
\label{tab:family-results}
\resizebox{\linewidth}{!}{%
\begin{tabular}{lccccc}
\toprule
Family & Pairs & Minimal & Top-5 & H2-free & H28-free \\
\midrule
\texttt{-EncodedCommand} & 7  & \texttt{-1.41 / 0.00} & \texttt{-0.99 / 0.00} & \texttt{-1.27 / 0.00} & \texttt{-1.11 / 0.00} \\
\texttt{DownloadFile}    & 146 & \texttt{-4.84 / 0.20} & \texttt{-5.07 / 0.15} & \texttt{-5.05 / 0.16} & \texttt{-4.41 / 0.07} \\
\texttt{DownloadString}  & 43  & \texttt{-3.77 / 0.44} & \texttt{-4.05 / 0.44} & \texttt{-4.12 / 0.51} & \texttt{-3.56 / 0.30} \\
\texttt{FromBase64String}& 35  & \texttt{-1.19 / 0.37} & \texttt{-1.26 / 0.34} & \texttt{-1.29 / 0.34} & \texttt{-1.09 / 0.34} \\
\texttt{IEX}             & 25  & \texttt{-3.03 / 0.32} & \texttt{-3.44 / 0.32} & \texttt{-3.48 / 0.44} & \texttt{-2.83 / 0.16} \\
\texttt{Invoke-Expression}& 18 & \texttt{-1.46 / 0.50} & \texttt{-1.61 / 0.50} & \texttt{-1.66 / 0.56} & \texttt{-1.34 / 0.44} \\
\texttt{Invoke-WebRequest}& 19 & \texttt{-1.92 / 0.47} & \texttt{-1.88 / 0.42} & \texttt{-1.86 / 0.47} & \texttt{-1.76 / 0.42} \\
\bottomrule
\end{tabular}
}
\end{table}

\subsection{Comparative Model Analysis}
Foundation-Sec and Llama-3.1-8B-Instruct share the same 32-layer, 32-head Llama-3.1 architecture, so the comparison can separate structural inheritance from fine-tuning-specific reweighting. The main comparative result is that the same basic circuit skeleton transfers: in Llama, \texttt{L0H11} is again the most recurrent early detector, appearing in \(14/18\) discovery pairs, and the strongest late integration layer is again Layer 12. What changes is the weighting of the late carrier. Foundation-Sec's route is centered on \texttt{L12H15/L12H5/L12H4/L12H28}, while Llama's traced route is centered on \texttt{L12H28/L12H5/L12H4/L12H13}. This pattern supports the interpretation that cybersecurity fine-tuning concentrated and semantically weighted a pre-existing structural circuit rather than creating a new PowerShell detector from scratch.

\begin{figure}[!htbp]
\centering
\includegraphics[width=0.92\linewidth]{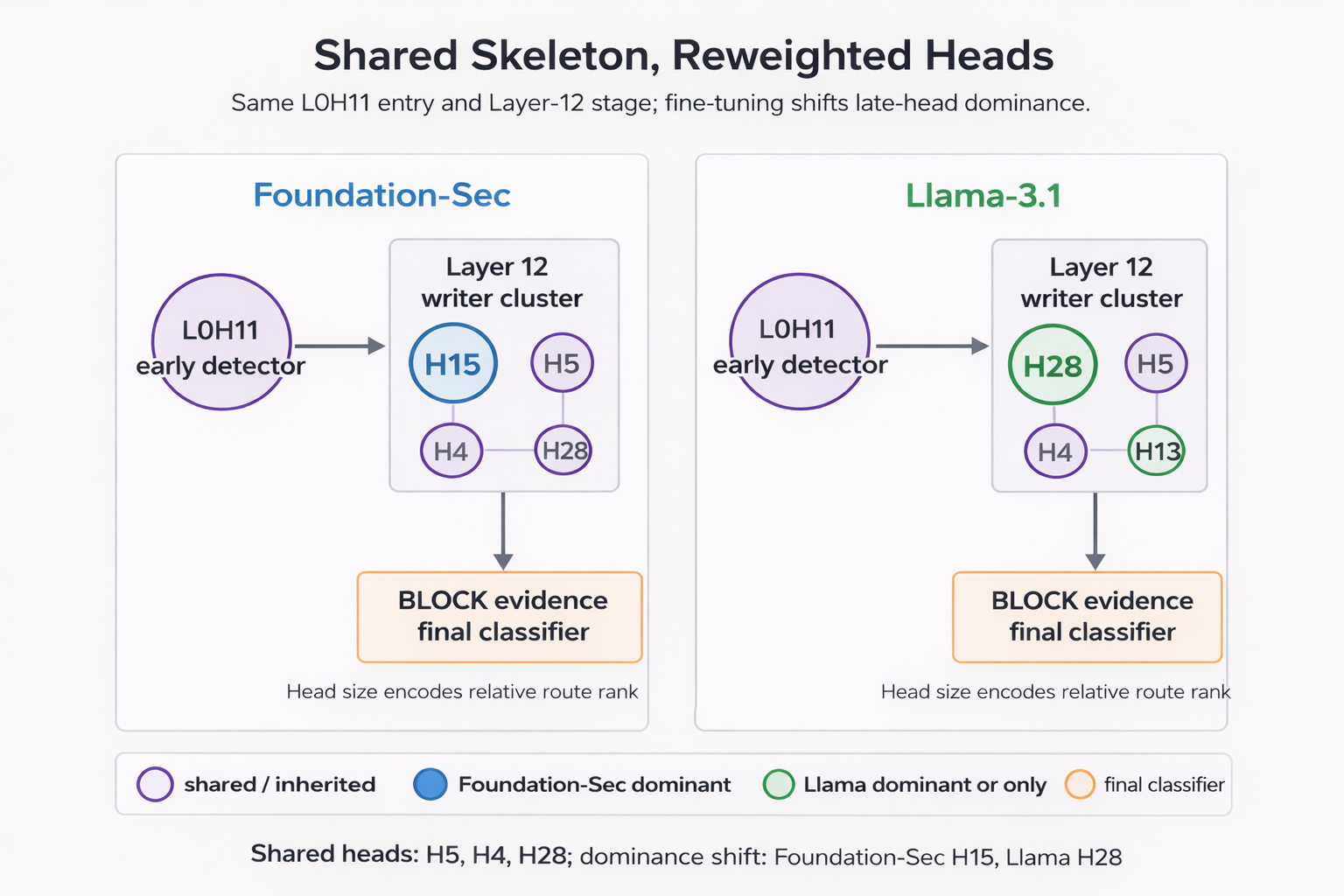}
\vspace{0.35em}
\includegraphics[width=0.92\linewidth]{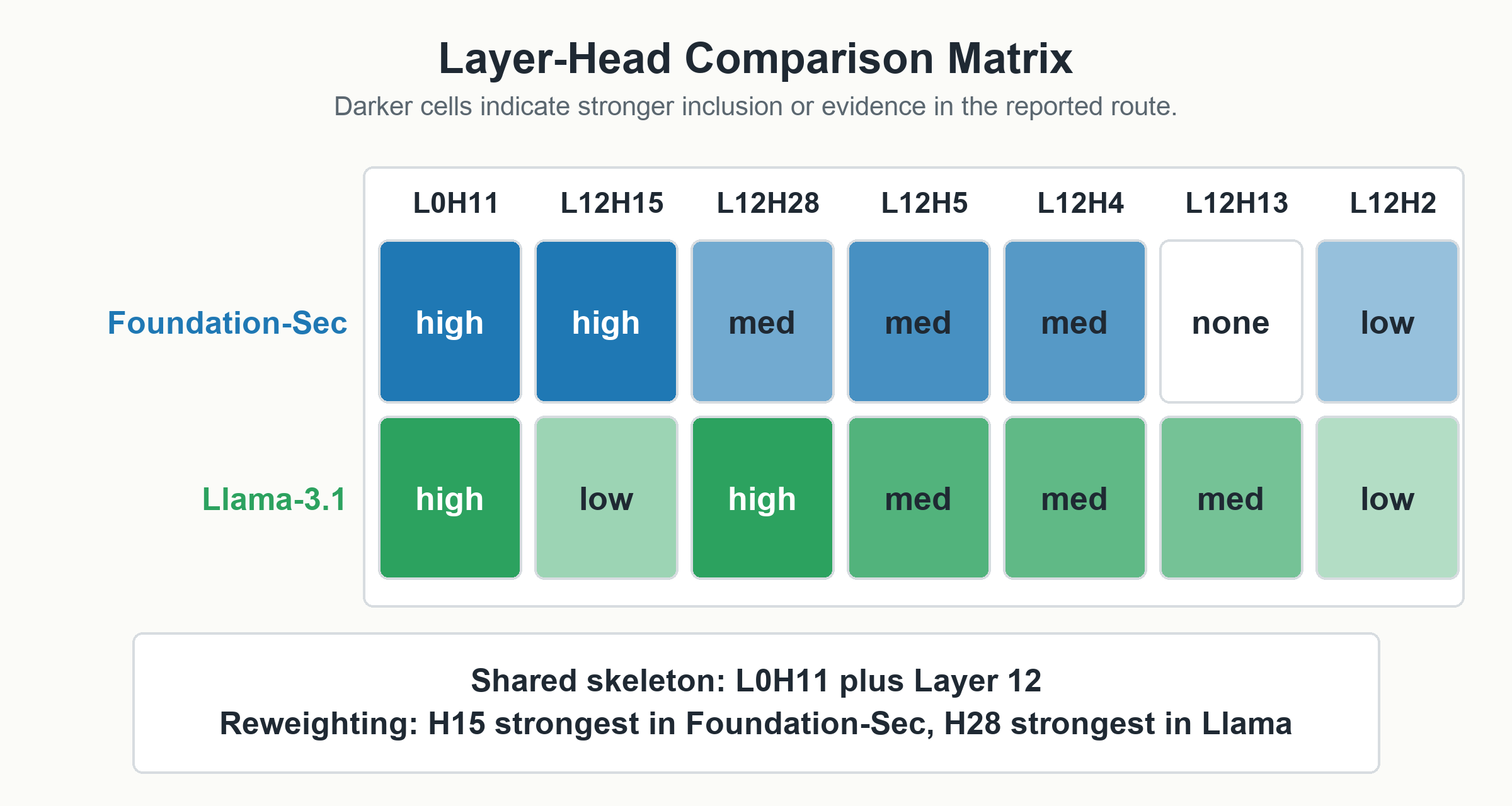}
\caption{Cross-model circuit summary. Foundation-Sec and Llama share the broad PowerShell classification skeleton, including the early detector and late integration layer, but fine-tuning shifts which late heads dominate and adds indicator-token semantics to the final classification signal.}
\label{fig:comparative-circuit-summary}
\end{figure}

Table~\ref{tab:model-comparison} quantifies the same pattern. The cleanest account is not that security training created the \texttt{L0H11}-to-Layer-12 route. Instead, fine-tuning concentrated causal weight into a smaller late-head set and added semantic associations between full-form PowerShell indicators and the final \texttt{BLOCK} decision, consistent with fine-tuning work showing node-level circuit persistence alongside edge and weighting changes \citep{wang2025finetuningcircuits}.
\begin{table}[!htbp]
\centering
\caption{Cross-model circuit comparison. Llama uses the same early detector and late writer layer as Foundation-Sec, but its strongest route is more diffuse and shifts late dominance toward \texttt{L12H28}.}
\label{tab:model-comparison}
\begin{tabular}{@{}p{0.20\linewidth}p{0.24\linewidth}p{0.22\linewidth}p{0.26\linewidth}@{}}
\toprule
Metric & Foundation-Sec-8B-Instruct & Llama-3.1-8B-Instruct & Interpretation \\
\midrule
Early detector & \texttt{L0H11} & \texttt{L0H11} & Shared structural entry point \\
Late writer layer & Layer 12 & Layer 12 & Shared late integration stage \\
Top late writer & \texttt{L12H15} & \texttt{L12H28} & Fine-tuning shifts late-head dominance \\
Llama \texttt{L0H11} recurrence & not measured & \(14/18\) discovery pairs & Early detector is not Foundation-Sec-specific \\
Minimal route patching & \(24/74\) flips on matched subset & \(-4.865\), \(11/74\) flips & Llama route is sufficient but less compact \\
Stronger route patching & \(\sim25/74\) flips on matched subset & \(-5.451\), \(18/74\) flips & Same layer, shifted carrier \\
Diffuse bundle patching & \(\sim25/74\) flips on matched subset & \(-6.158\), \(36/74\) flips & Llama needs a broader head bundle \\
\bottomrule
\end{tabular}
\end{table}

The model-level accuracy comparison points in the same direction. On the matched 74-pair adversarial-prompt baseline, Foundation-Sec gains \(4.7\) percentage points over Llama (\(100\%\) vs.\ \(95.3\%\)), driven by benign accuracy rather than malicious recall. But the causal margins do not simply get stronger: Llama's late carrier patching effect is larger than Foundation-Sec's on a comparable route (\(-5.45\) vs.\ \(-3.82\)), and Llama's mean malicious logit margin is also higher (\(8.15\) vs.\ \(3.75\)). The resulting picture is that fine-tuning preserves an inherited structural route, concentrates how it is used, and adds MLP-mediated semantic weighting before the classification boundary. That semantic layer helps baseline classification but, as the evasion benchmark shows next, creates attack surfaces when aliases, reconstruction, indirection, or case mutation change the indicator surface.

\subsection{Evasion Results}
Foundation-Sec's results separate cleanly by tier. Because the benchmark is conditioned on baseline-correct malicious seeds, these evasion failures should be read as misses on scripts the model initially handled correctly before rewriting. In \texttt{direct\_v1}, Foundation-Sec stays correct on \(38/44\) accepted variants and the misses are concentrated in two techniques rather than spread broadly across the conservative benchmark. The strongest direct-tier failure remains simple alias substitution in the \texttt{Invoke-WebRequest} family: replacing the full command name with \texttt{iwr} yields misses on all four accepted variants tested. The only other positive result in that tier is narrower. Indirect \texttt{DownloadString} method invocation produces misses in two of six variants, and both of those misses come from \texttt{FromBase64String}-linked seeds rather than from the direct \texttt{DownloadString} seeds themselves.

Foundation-Sec classification accuracy is still strong under the second tier. In \texttt{reconstructive\_v2}, it stays correct on \(42/46\) candidates. All four misses come from one keyword-hiding technique, \texttt{invoke\_expression\_format\_string}. The other reconstructive rewrites, including backtick splitting, Base64 or ASCII reconstruction, subexpression strings, and zero-width normalization for \texttt{-EncodedCommand}, remain negative in the current run. Across the original two tiers together, Foundation-Sec stays correct on \(80/90\) accepted variants, so the observed weaknesses are specific and technique-concentrated rather than benchmark-wide.

\texttt{case\_mutation\_v3} strengthens the same interpretation and broadens the applicable miss class. Under the raw prompt, Foundation-Sec produces \(9/63\) misses on accepted case-mutation variants, while Llama produces \(0/63\) misses on the same variants. The failures come from three groups: \texttt{invoke\_expression\_alternating\_case} produces \(4/4\) misses, \texttt{invoke\_expression\_lowercase} produces \(1/4\) miss, and case-mutated \texttt{IEX} alias variants produce \(4/4\) misses. The remaining accepted case-mutation variants, including structural \texttt{IEX} rewrites, leave substantial positive residual margins. Because the positive groups are still small and targeted, these results should be read as localized counterexamples and red-teaming priorities, not as stable estimates of real-world evasion prevalence.

The benchmark is therefore not a broad obfuscation leaderboard. It is a controlled seed/variant attribution harness for red-team prioritization: each accepted miss preserves the relevant behavior while isolating which syntactic transformation changed the model's decision.

\begin{table}[!htbp]
\centering
\caption{Foundation-Sec evasion benchmark status. Variants satisfy the parse or manual-validation checks and behavioral invariants from the implemented pipeline.}
\label{tab:evasion-results}
\begin{tabular}{@{}p{0.03\linewidth}p{0.43\linewidth}p{0.23\linewidth}p{0.05\linewidth}p{0.15\linewidth}@{}}
\toprule
Tier & Technique & Family & Pairs & Outcome \\
\midrule
\texttt{v1}& \texttt{invoke\_webrequest\_alias} & \texttt{keyword\_hiding} & 4 & $4/4$ misses \\
\texttt{v1}& \texttt{downloadstring\_psobject\_invoke} & \texttt{string\_construction} & 6 & $2/6$ misses \\
\texttt{v1}& all other techniques & mixed & 34 & $0/34$ misses \\
\texttt{v2}& \texttt{invoke\_expression\_format\_string} & \texttt{keyword\_hiding} & 4 & $4/4$ misses \\
\texttt{v2}& all other techniques & mixed & 42 & $0/42$ misses \\
\texttt{v3}& \texttt{invoke\_expression\_alternating\_case} & \texttt{case\_mutation} & 4 & $4/4$ misses \\
\texttt{v3}& \texttt{invoke\_expression\_lowercase} & \texttt{case\_mutation} & 4 & $1/4$ misses \\
\texttt{v3}& case-mutated \texttt{IEX} alias & \texttt{case\_mutation} & 4 & $4/4$ misses \\
\texttt{v3}& all other accepted case variants & mixed & 51 & $0/51$ misses \\
\bottomrule
\end{tabular}
\end{table}

\begin{table}[!htbp]
\centering
\caption{Cross-model evasion comparison under the controlled raw-prompt classifier framing. Llama has no misses on the same evaluated variants, while Foundation-Sec's misses indicate a prompt-sensitive semantic layer added on top of the inherited structural circuit.}
\label{tab:cross-model-evasion}
\begin{tabular}{llll}
\toprule
Model & Benchmark tier & Variants & Misses \\
\midrule
Llama-3.1-8B-Instruct & \texttt{direct\_v1} & 44 & \(0/44\) \\
Foundation-Sec-8B-Instruct & \texttt{direct\_v1} & 44 & \(6/44\) \\
Llama-3.1-8B-Instruct & \texttt{reconstructive\_v2} & 46 & \(0/46\) \\
Foundation-Sec-8B-Instruct & \texttt{reconstructive\_v2} & 46 & \(4/46\) \\
Llama-3.1-8B-Instruct & \texttt{case\_mutation\_v3} & 63 & \(0/63\) \\
Foundation-Sec-8B-Instruct & \texttt{case\_mutation\_v3} & 63 & \(9/63\) \\
\bottomrule
\end{tabular}
\end{table}

The Llama result is stable across the evaluated prompt framings: it produces \(0/44\) misses on \texttt{direct\_v1}, \(0/46\) misses on \texttt{reconstructive\_v2}, and \(0/63\) misses on \texttt{case\_mutation\_v3}. Foundation-Sec's failure surface is more prompt-sensitive. Under the raw prompt, Foundation-Sec's original-tier misses clustered around techniques that removed specific literal indicator tokens: \texttt{invoke\_webrequest\_alias} (\(4/4\) misses) and \texttt{downloadstring\_psobject\_invoke} (\(2/6\)). Under the adversarial prompt (``classify based on overall purpose, not individual constructs''), \texttt{invoke\_webrequest\_alias} was fully fixed: the model shifted to a more holistic assessment and recovered the malicious intent through other signals. But a new set of misses simultaneously appeared on \texttt{Invoke-Expression} obfuscation variants, and \texttt{case\_mutation\_v3} shows the same pattern more sharply. On the 43 non-\texttt{IEX} case-mutation candidates, the adversarial prompt increases Foundation-Sec misses from \(5/43\) to \(12/43\), retaining \(4/4\) misses on \texttt{invoke\_expression\_alternating\_case} and adding \(4/4\) misses on \texttt{invoke\_expression\_lowercase}, \(2/4\) on \texttt{downloadstring\_lowercase}, and \(2/6\) on \texttt{downloadstring\_alternating\_case}. This supports the semantic-layer interpretation: fine-tuning added useful associations around canonical indicators such as \texttt{Invoke-Expression} and \texttt{Invoke-WebRequest}, but those associations can be bypassed or reshaped by aliases, command reconstruction, string construction, execution indirection, and case mutation. The adversarial-prompt comparison also suggests that the inversion itself is Foundation-Sec-specific rather than a generic Llama property: on the same adversarial scripts, Llama's Layer 13 residual stays positive, consistent with the base model lacking the same fine-tuned indicator-token associations.

\begin{table}[!htbp]
\centering
\caption{Representative side-by-side examples of successful evasion from both benchmark tiers. In \texttt{direct\_v1}, replacing the explicit \texttt{Invoke-WebRequest} token with the alias \texttt{iwr} is sufficient to flip Foundation-Sec's prediction. In \texttt{reconstructive\_v2}, a format-string reconstruction of \texttt{Invoke-Expression} likewise produces a Foundation-Sec miss while preserving the surrounding payload logic.}
\label{tab:evasion-example}
\begin{tabular}{p{0.14\linewidth} p{0.14\linewidth} p{0.64\linewidth}}
\toprule
Tier & Version & Script fragment \\
\midrule
\texttt{v1}& Seed &
\texttt{Invoke-WebRequest -Uri http://stderr.pl/procdump.exe -OutFile c:\textbackslash temp\textbackslash sendme.exe} \\
\texttt{v1}& Variant &
\texttt{iwr -Uri http://stderr.pl/procdump.exe -OutFile c:\textbackslash temp\textbackslash sendme.exe} \\
\midrule
\texttt{v2}& Seed &
\texttt{Invoke-Expression \$(New-Object IO.StreamReader (...)).ReadToEnd();} \\
\texttt{v2}& Variant &
\texttt{\&(('{0}{1}' -f 'Invoke-','Expression')) \$(New-Object IO.StreamReader (...)).ReadToEnd();} \\
\bottomrule
\end{tabular}
\end{table}
The key mechanistic finding on the successful evasion cases is that the internal classification signal does not simply disappear when the indicator transformation is applied. Under \texttt{invoke\_webrequest\_alias}, the late-layer attention heads that normally carry malicious evidence still fire on the miss variants, so the internal evidence is still there. What changes is how feed-forward computation before the Layer 13 boundary uses that evidence: it reverses the signal, turning what was a push toward malicious into a push toward benign. The evasion is therefore better understood as signal reversal by MLP-side computation near the classification boundary than as deletion of the detection mechanism. The \texttt{invoke\_expression\_format\_string} misses show the same pattern: the early and late attention components still respond to the malicious content (\(4/4\) variants flip under patching), but MLP layers 0--12 introduce a competing signal before the Layer 13 boundary that overrides their output. The case-mutation results further narrow the failure mode: alternating-case \texttt{Invoke-Expression} and \texttt{IEX} preserve command identity, position, and arguments, so their misses point to token-embedding sensitivity rather than syntactic or semantic invalidity. Because Foundation-Sec and Llama share the same tokenizer, the cross-model difference cannot be explained by token segmentation alone; it must come from how fine-tuning changed the downstream MLP transformations of those token embeddings. These cases are consistent with the observation that fine-tuning added indicator-token semantic associations that interact with the inherited detection circuit in ways that can invert the final output when those indicators are transformed or absent \citep{lindsey2025biology}.

The current evidence does \emph{not} support a generic statement that direct or reconstructive obfuscation automatically defeats these models. In Foundation-Sec \texttt{direct\_v1}, misses occur only in \texttt{invoke\_webrequest\_alias} and the narrower \texttt{downloadstring\_psobject\_invoke} pattern concentrated in \texttt{FromBase64String}-linked seeds. In Foundation-Sec \texttt{reconstructive\_v2}, misses occur only in \texttt{invoke\_expression\_format\_string}; the other tested backtick, Base64 or ASCII, subexpression, and zero-width-normalization techniques remain robust among the accepted variants in the current run. In \texttt{case\_mutation\_v3}, misses concentrate in \texttt{Invoke-Expression} and the \texttt{IEX} alias, while DownloadFile, most DownloadString variants, and Invoke-WebRequest case variants remain robust or near-miss. Llama's controlled raw-prompt results, \(0/44\) misses on \texttt{direct\_v1}, \(0/46\) misses on \texttt{reconstructive\_v2}, and \(0/63\) misses on accepted case-mutation variants, show that the same architectural circuit can be less brittle when it remains more diffuse and less tied to Foundation-Sec's full-form indicator semantics.

\begin{figure}[!htbp]
\centering
\includegraphics[width=\linewidth]{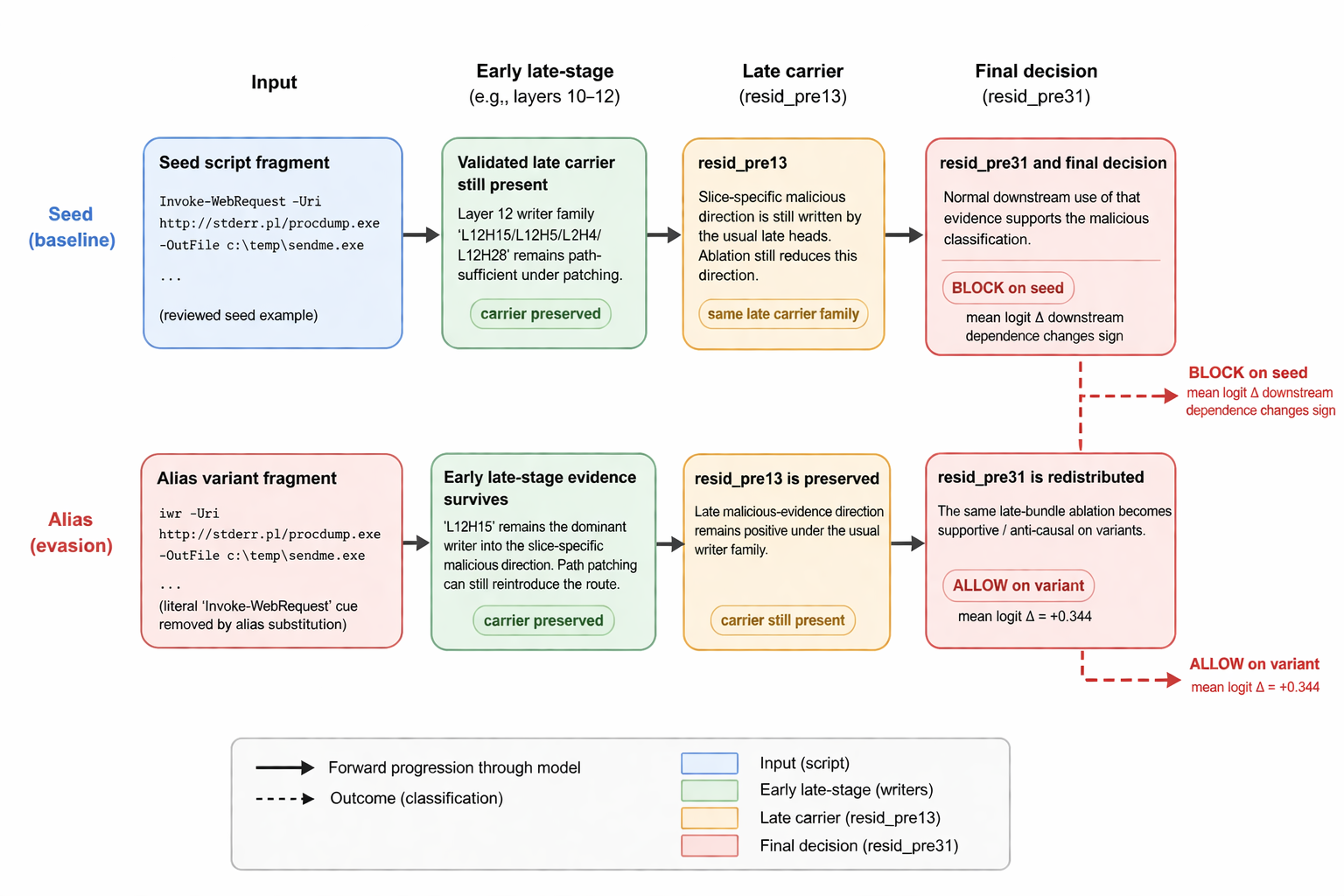}
\caption{Indicator-transformation evasion flowchart for the strongest current failure modes. The attack preserves malicious behavior while replacing or reconstructing a canonical indicator; the inherited malicious-evidence signal remains internally available, but fine-tuned feed-forward computation near the classification boundary reverses or misuses that signal before final classification.}
\label{fig:alias-evasion-flowchart}
\end{figure}

\subsection{Monitoring Fine-Tuning Drift}
The comparative and evasion results support a practical operational extension. Foundation-Sec inherits the underlying PowerShell classification route from Llama, but its failures after fine-tuning are not best explained as deletion of that inherited mechanism. Instead, the current evidence points to a narrower change in how fine-tuned feed-forward computation uses that inherited signal before the Layer 13 boundary: the relevant malicious-evidence representation still appears at the causally important \texttt{resid\_pre13} site, while MLP-side computation can suppress, redirect, or invert its effect for particular semantic families. This makes \texttt{resid\_pre13} a useful monitoring anchor for family-level regression testing after security fine-tuning.

\paragraph{Linear probe at the circuit boundary.}
We fit a linear probe (Ridge regression, 5-fold cross-validation) from the \texttt{resid\_pre13} activation of each malicious script to the per-head contribution of the four core Layer-12 heads measured by the head trace experiment. This use of a cheap readout from a causally localized site follows the broader circuit-discovery pattern of using patching results to identify important activations, then using cheaper approximations or readouts to scale analysis across examples \citep{conmy2023automated,nanda2023attributionpatching}. The probe achieves cross-validated \(r = 0.80\)--\(0.87\) for \texttt{L12H4}, \texttt{L12H5}, \texttt{L12H15}, and \texttt{L12H28} (\(n = 206\), all \(p < 10^{-45}\)). These values confirm that each head's contribution to the contrastive direction is \emph{linearly} encoded in \texttt{resid\_pre13}, meaning the information is directly readable as a simple linear projection of the inherited representation, not just extractable by a more flexible model.

Operationally, this gives a cheap drift quantity: run one forward pass through Llama and one through Foundation-Sec on canonical clean inputs, apply the same learned linear projection to each model's \texttt{resid\_pre13} activations, and compare the projected head contribution by capability family. If Foundation-Sec has preserved the inherited representation for a family, the family mean should remain close to the Llama mean. If fine-tuning compresses, rotates, or inverts that family's representation at the circuit boundary, the projected contribution will diverge. Applied cross-model at the script level, the same probe predicts Foundation-Sec's causal circuit dependence (\texttt{delta\_logit\_diff} under top-5 L12 ablation) at \(r = 0.41\) (\(n = 293\), \(p < 10^{-12}\)). This is useful but also sets the ceiling: per-script prioritization is noisy, so the intended operational unit is a family-level red-teaming priority, where averaging over scripts stabilizes the signal.
\begin{figure}[!htbp]
\centering
\includegraphics[width=0.92\linewidth]{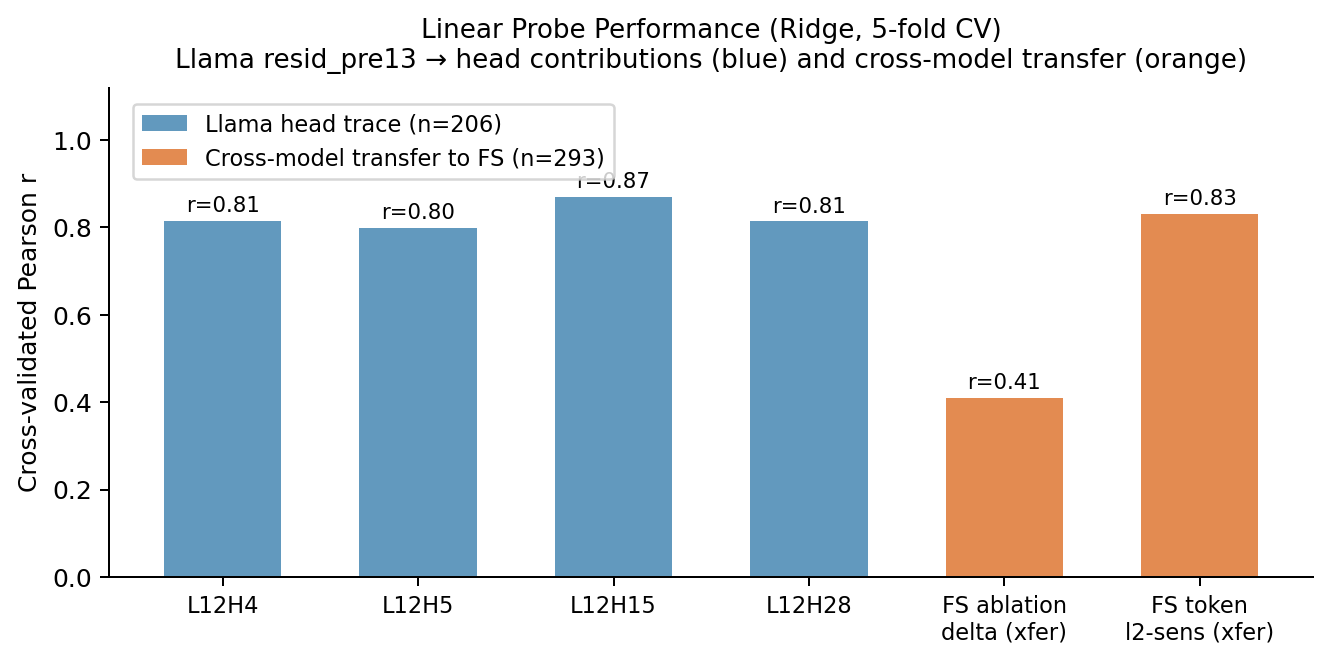}
\caption{Cross-validated linear probe performance (Ridge regression, 5-fold CV). Blue bars show how well Llama's \texttt{resid\_pre13} activation predicts each Layer-12 head's contribution to the contrastive direction (\(r = 0.80\)--\(0.87\), \(n = 206\)). Orange bars show cross-model transfer: the same Llama activations predicting Foundation-Sec's circuit dependence under top-5 ablation (\(r = 0.41\)) and indicator-token L2 sensitivity (\(r = 0.83\)). The high within-model r values confirm each head's contribution is linearly encoded at the circuit boundary; the lower ablation-delta transfer (\(r = 0.41\)) reflects the ceiling imposed by the cross-model fine-tuning gap.}
\label{fig:probe-r-values}
\end{figure}

\begin{figure}[!htbp]
\centering
\includegraphics[width=0.92\linewidth]{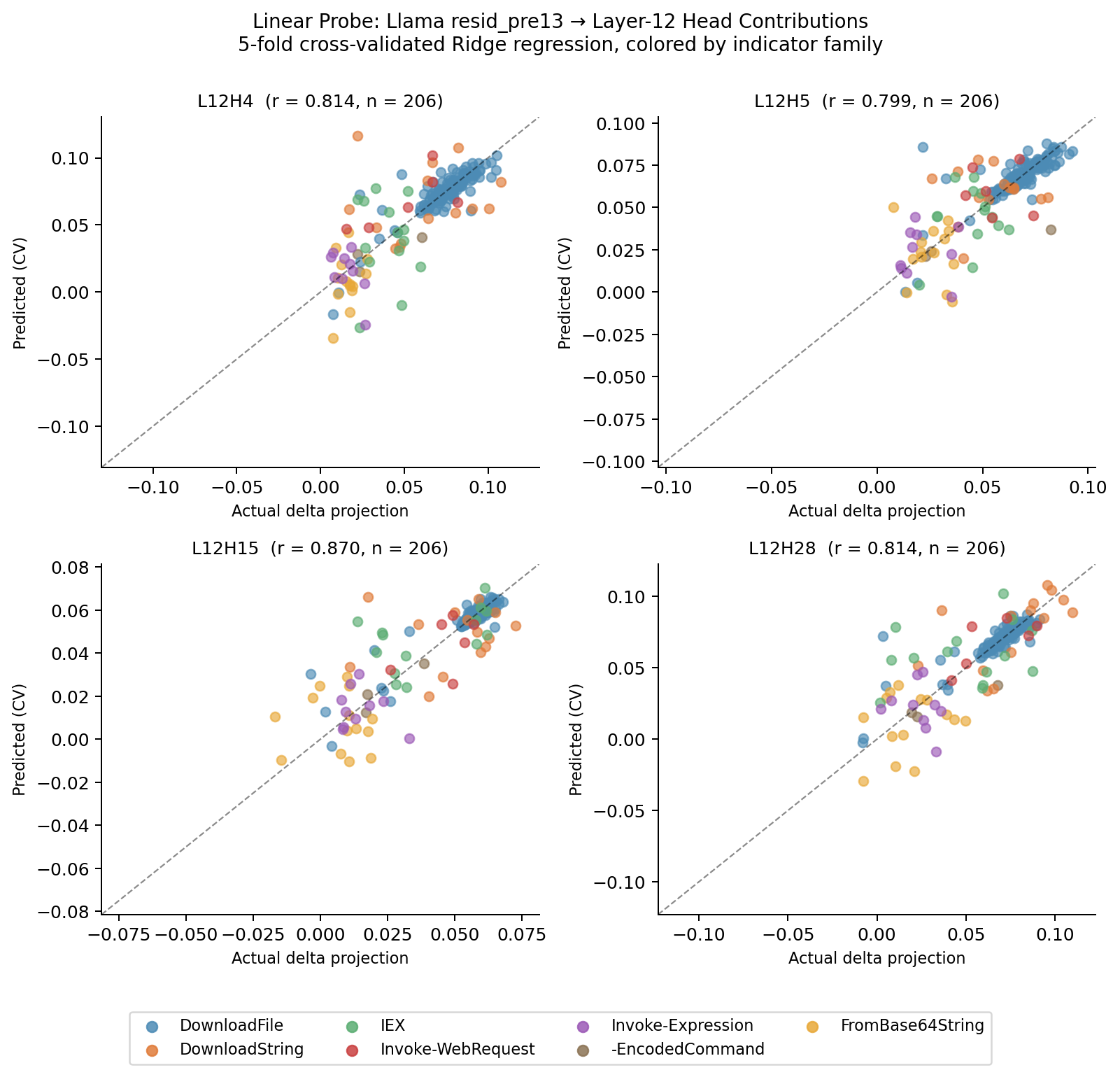}
\caption{Scatter plots of predicted versus actual Layer-12 head contributions for each of the four core heads (\texttt{L12H4}, \texttt{L12H5}, \texttt{L12H15}, \texttt{L12H28}). Each point is one malicious script; the dashed diagonal is the identity line. Colors indicate indicator family. The tight alignment around the diagonal across all four panels confirms that the probe's \(r = 0.80\)--\(0.87\) performance is consistent and not driven by outliers or a single family.}
\label{fig:probe-scatter}
\end{figure}

The practical implication is not that the probe reliably ranks individual scripts by future evasion risk. Rather, the family-level average of the projected contribution under Llama versus Foundation-Sec reveals which capability families changed most in how strongly the inherited circuit is represented after fine-tuning. Those families are the natural first targets for adversarial variant generation.

\paragraph{Indicator-token sign inversion as an evasion signal.}
The linear probe's cross-model transfer is strongest at the family level, where per-script noise averages out. But projection magnitude alone does not determine which changed families are specifically vulnerable to behavior-preserving indicator transformations. A sharper signal comes from directly measuring how much each model's malicious classification confidence changes when canonical indicator tokens are removed. For each malicious script, we replace the indicator-token embeddings with an average neutral embedding (the mean across the model's full vocabulary) and measure the resulting change in the model's confidence score.

\begin{figure}[!htbp]
\centering
\includegraphics[width=0.92\linewidth]{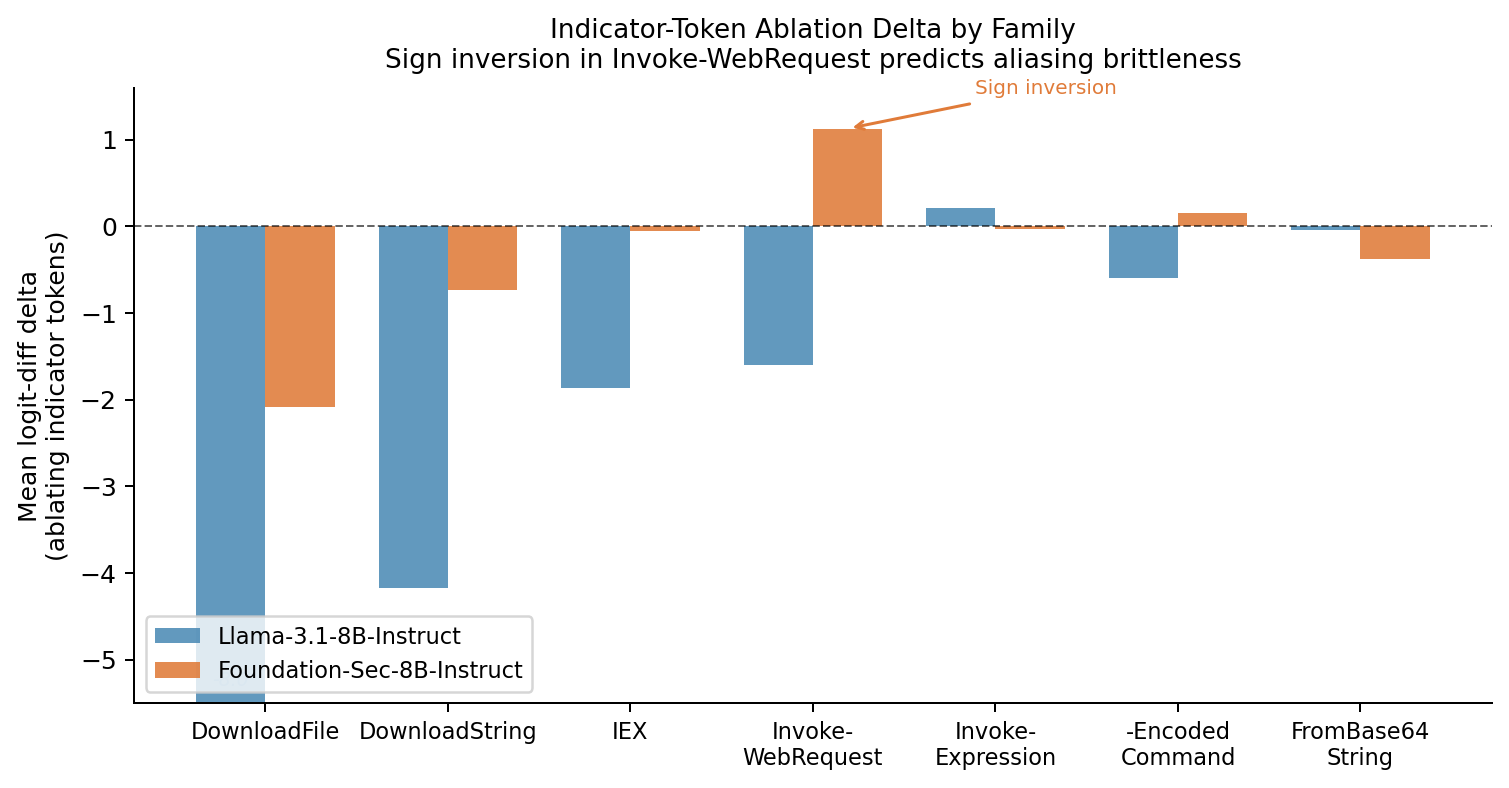}
\caption{Mean logit-diff delta when indicator tokens are mean-ablated, broken down by indicator family and model. A negative delta means removing the indicator tokens reduces malicious confidence (the token was acting as a driver of the malicious prediction). A positive delta means removing them increases malicious confidence (the token was acting as a suppressor). \texttt{Invoke-WebRequest} shows the clearest divergence: Foundation-Sec (orange) has a large positive delta (\(+1.13\)), while Llama (blue) shows a strong negative delta (\(-1.60\)). This divergence identifies \texttt{Invoke-WebRequest} as the family where the canonical indicator changed role between models and where alias substitution (\texttt{iwr}) produces consistent classification failures. The DownloadFile Llama bar extends below the axis limit (value: \(-8.0\)); the y-axis is capped at \(-5.5\) to keep the smaller-delta families readable.}
\label{fig:ablation-delta}
\end{figure}

Most families in Foundation-Sec show the expected pattern: removing the indicator tokens reduces malicious confidence, consistent with those tokens acting as drivers of the malicious prediction. \texttt{Invoke-WebRequest} is the clear exception. Ablating its canonical command tokens \emph{increases} average malicious confidence (\(+1.13\) mean logit-diff delta; 73.7\% of scripts show a positive delta), while Llama shows a strong negative delta (\(-1.60\)) for the same family. The divergence between the two models is most pronounced here: all other families stay within a narrow band, while \texttt{Invoke-WebRequest} is the only family with a large positive FS delta alongside a clearly negative Llama delta. Figure~\ref{fig:ablation-delta} shows the full family-level comparison.

The mechanistic interpretation is that fine-tuning appears to have led Foundation-Sec to classify \texttt{Invoke-WebRequest} scripts through URL patterns, payload structure, and command arguments rather than through the canonical command token itself. The evidence separates two effects. On canonical inputs, mean-ablating the full command token increases malicious confidence, so the token itself behaves like a suppressor. On the successful \texttt{iwr} variants, however, the validated late bundle still appears while feed-forward computation before the Layer 13 boundary changes how that evidence is used, indicating that the alias transformation also disrupts or reroutes the command-anchored payload-context signal. The miss therefore does not follow from suppressor removal alone; it follows from suppressor removal plus loss or inversion of the stronger contextual evidence path, with the latter dominating in the accepted variants tested here.

The linear probe also quantifies how much a script's internal representation shifts when its indicator tokens are removed: this shift is linearly predictable from the base-model activation at \(r = 0.83\) (\(n = 293\), \(p < 10^{-75}\)), confirming that token dependence is a stable, readable property of the inherited representation. The direction of that dependence, whether removing the tokens increases or decreases malicious confidence, is the operative transformation-sensitivity signal. A family where the fine-tuned model's ablation delta flips sign relative to the base model should be prioritized for behavior-preserving indicator transformations, and this flip is detectable on canonical clean inputs without generating any evasion variants in advance.

The output is not a prediction for individual scripts; it is a ranked list of command families to red-team. Operationally, this motivates a two-stage family-level monitoring recipe:
\begin{enumerate}[leftmargin=1.5em,itemsep=0.2em,topsep=0.2em,parsep=0pt]
  \item \textbf{Fit a drift detector:} using the base model and a set of canonical malicious scripts, train a linear probe that predicts how strongly each key attention head contributes to the classification decision from the model's internal state at the circuit boundary.
  \item \textbf{Compare base and fine-tuned models:} run both models on the same canonical inputs, apply the probe to each, and compare the results by command family. Families where the fine-tuned model's score diverges strongly from the base model have drifted most in how they use the inherited circuit.
  \item \textbf{Run the sign test:} for each family, remove the indicator tokens in both models (replacing them with a neutral average embedding) and measure whether confidence goes up or down. A family where the fine-tuned model's confidence \emph{increases} when the indicator is removed, while the base model's confidence \emph{decreases}, has undergone a role reversal and is a transformation-sensitive attack surface.
  \item \textbf{Prioritize red-teaming:} generate alias, command-reconstruction, string-construction, execution-indirection, and case-mutation variants for the flagged families and test them against the fine-tuned model.
\end{enumerate}

Both steps use only canonical clean inputs. No evasion variants need to be generated in advance and no retraining is required. Full circuit discovery is useful for choosing a high-quality monitoring site, as in our use of \texttt{resid\_pre13}, but the indicator-token sign test itself only requires comparing base and fine-tuned responses to canonical indicator ablations. In this study, that sign test correctly identifies \texttt{Invoke-WebRequest} as the red-teaming priority before any alias variants are tested.

\section{Future Work}
Next steps for this line of work can include extending the validation cohort to cover additional indicator families beyond the seven reported here, additional security fine-tunes beyond Foundation-Sec, and deeper per-family pair counts, so that the current 293-pair model-pair result can be stress-tested with a more independent validation design. The current strongest evasion groups are compact but consistent: \(4/4\) accepted \texttt{Invoke-WebRequest} alias variants, \(4/4\) accepted \texttt{Invoke-Expression} format-string variants, \(4/4\) \texttt{Invoke-Expression} alternating-case variants, and \(4/4\) accepted case-mutated \texttt{IEX} alias variants. Expanding each group would turn the present red-teaming signals into stronger population-level estimates.

On the robustness side, the evasion benchmark can expand along two fronts: deeper coverage inside the already active \texttt{keyword\_hiding}, \texttt{string\_construction}, \texttt{execution\_indirection}, and \texttt{case\_mutation} families, and concrete tests for two currently theoretical families. \texttt{network\_object\_indirection} would hide direct networking surfaces behind equivalent object-construction, method-dispatch, or reflection patterns. \texttt{staged\_payload\_flow} would preserve the same malicious behavior while splitting retrieval, decoding, allocation, and execution across intermediate variables or helper functions. A broader cohort could also relax the current baseline-correct matched-pair filter and measure whether the same mechanisms persist outside the causally clean subset used here.

On the mechanistic side, the current evidence points to a distributed MLP contribution across Layers 0--12 rather than a single feed-forward component that explains the reversal by itself. Future work should characterize how that collective MLP effect accumulates before the Layer 13 boundary and how it interacts with the surviving late-attention evidence. 

\section{Conclusion}
Security fine-tuning improves LLM classification on the inputs it was trained to handle and, as this study shows, can simultaneously introduce evasion vulnerabilities that standard evaluation is not designed to detect. A model trained on a corpus where specific indicator tokens are correlated with maliciousness will score well on held-out test data while silently failing on scripts that preserve behavior but transform those indicators. This is not a model quality problem in the conventional sense; it is a gap between what evaluation measures and what adversaries can exploit.

We demonstrate this concretely for Foundation-Sec-8B-Instruct using mechanistic interpretability tools applied to a natural base/fine-tuned model pair. Causal interventions localize the classification circuit to a Layer-12 attention bundle inherited from Llama. Fine-tuning concentrated and semantically specialized this existing structure rather than building a new one. That specialization is what creates the vulnerability: Foundation-Sec learned indicator semantics that help baseline classification behavior but can be disrupted by behavior-preserving transformations. The evasion benchmark confirms the consequence: consistent misses on \texttt{iwr} substitution, \texttt{Invoke-Expression} reconstruction, and case-mutated \texttt{Invoke-Expression}/\texttt{IEX} that Llama, with its more diffuse inherited representation, does not share on the same evaluated variants. To our knowledge this is among the first mechanistic circuit analyses in a cybersecurity LLM context, combining causal validation, cross-model comparison, and a matched seed/variant evasion benchmark \citep{choi2020malicious,garciacarrasco2024detecting}.

Taken together, these findings show that security fine-tuning can create an MLP-mediated semantic layer that both helps baseline classification behavior and creates transformation-, case-, and prompt-sensitive attack surfaces.

The practical output is a two-stage pre-deployment workflow that any team fine-tuning an LLM for security classification can apply. First, a linear probe from base-model activations at the classification boundary (\(r = 0.80\)--\(0.87\) per head) identifies which capability families drifted most in how strongly the inherited circuit engages after fine-tuning. Second, an indicator-token sign test compares how each model's confidence responds to indicator-token ablation on canonical clean scripts and flags the specific families where the token role flipped from driver to suppressor. Both steps require only forward passes on canonical inputs; no evasion variants need to be generated in advance. In this study the sign test correctly identifies \texttt{Invoke-WebRequest} as the red-teaming priority before any alias variants are tested, while the broader benchmark also surfaces \texttt{Invoke-Expression} reconstruction and case-mutation misses. The method is not a per-script evasion predictor, but it provides a principled family-level prioritization signal that standard accuracy evaluation does not.

Overall, the practical recommendation is straightforward: treat fine-tuning as a potential source of representation drift. Before deployment, compare the base and fine-tuned models on canonical inputs, identify which command families changed most, and red-team those families with behavior-preserving variants. The goal is not to predict every evasion. The goal is to find the parts of the task where fine-tuning may have made the model semantically brittle. 

\appendix

\section{Dataset and Evaluation Cohorts}
\label{app:dataset-cohorts}
The source corpus contains 2{,}556 labeled PowerShell scripts, derived from prior work \citep{FANG202130}. The class balance after filtering is approximately even. A substantial long tail of long scripts motivated a 12{,}000-character pre-processing cap when building the analysis manifests and matched cohorts, so benign and malicious scripts were paired and selected under explicit length control rather than letting the very longest scripts dominate the mechanistic runs.

The main evidentiary basis for this draft is a 293-pair within-family matched cohort drawn from 456 unique scripts across the seven indicator families. In this cohort, benign and malicious scripts share suspicious indicators while remaining length-controlled enough for the intervention pipeline. The cohort was filtered from 380 candidate pairs by requiring both members of each pair to be correctly classified at baseline. This filter makes the intervention and evasion analyses easier to interpret: when a route is patched, ablated, or transformed, the measured effect is relative to examples both models initially handled correctly.

\section{Claim Ladder}
\label{app:claim-ladder}

\begin{table}[!htbp]
\centering
\small
\setlength{\tabcolsep}{4pt}
\renewcommand{\arraystretch}{0.95}
\caption{Claim ladder for the current study. This table separates discovery context from the main validated claim, cross-model comparison, and evasion evidence.}
\label{tab:claim-ladder}
\begin{tabular}{@{}p{0.11\linewidth}p{0.4\linewidth}p{0.47\linewidth}@{}}
\toprule
Claim level & Main statement & Role in this draft \\
\midrule
Discovery & Early Layer 0 detector family, especially \texttt{L0H11} and \texttt{L0H9} & Initial localization result motivating the narrower validated route, but not final evidence by itself \\
Validation & \texttt{L0H11 $\rightarrow$ L12H15/L12H5/L12H4} & Cleanest minimal direct branch on the 293-pair cohort \\
Validation & \texttt{L12H15/L12H5/L12H4/L12H28} & Stronger late-stage bundle on the 293-pair cohort \\
Refinement & \texttt{L12H2} helps some grouped ablations but weakens the main patching comparison & Auxiliary family-sensitive helper rather than part of the stable sufficiency core \\
Comparison & Llama shares \texttt{L0H11} and Layer 12 but shifts late dominance toward \texttt{L12H28} & Evidence that fine-tuning reshapes an inherited circuit rather than creating the structural route \\
Evasion & \texttt{direct\_v1}, \texttt{reconstructive\_v2}, and \texttt{case\_mutation\_v3} produce Foundation-Sec misses that Llama does not share on the same evaluated variants & Mechanistic reading of the current benchmark failures; supports signal reversal rather than deletion and broadens the attack surface beyond alias substitution \\
\bottomrule
\end{tabular}
\end{table}

\section*{Trademark Notice}
Llama is a trademark of Meta Platforms, Inc. PowerShell is a trademark of Microsoft Corporation. Cisco is a trademark or registered trademark of Cisco and/or its affiliates. GitHub is a trademark or registered trademark of GitHub, Inc. All other trademarks are the property of their respective owners.

\bibliographystyle{plainnat}
\IfFileExists{references.bib}{\bibliography{references}}{\bibliography{../references}}

\end{document}